\documentstyle[preprint,aps]{revtex}

\draft
\newcommand{\ba}{\begin{eqnarray}}
\newcommand{\ea}{\end{eqnarray}}
\begin{document}

\title{F-spin as a Partial Symmetry}
\author{ A. Leviatan$^{1,2,3}$ and J.N. Ginocchio$^{2,3}$}
\address{$^{1}$~Racah Institute of Physics, The Hebrew University,
Jerusalem 91904, Israel}
\address{$^{2}$~Theoretical Division, Los Alamos National Laboratory, Los
Alamos, New Mexico 87545}
\address{$^{3}$Institute for Nuclear Theory, University of Washington, 
Box 351550, Seattle, WA 98195}
%\date{\today}
\maketitle

\begin{abstract}
We use the empirical evidence that F-spin multiplets exist in nuclei for 
only selected states as an indication that F-spin can be regarded as 
a partial symmetry.
We show that there is a class of non-F-scalar IBM-2 
Hamiltonians with partial F-spin symmetry, which reproduce the known 
systematics of collective bands in nuclei.  
These Hamiltonians predict that the scissors states have good F-spin and 
form F-spin multiplets, which is supported by the existing data.  
\end{abstract}
\vspace{12pt}

The interacting boson model (IBM-2) \cite{iac87,arima77,otsuka78} 
describes collective low-lying states in even-even nuclei in terms of 
monople ($s_{\rho}$) and quadrupole ($d_{\rho}$) proton ($\rho=\pi$) 
and neutron ($\rho=\nu$) bosons.
Microscopic, shell-model-based interpretation of the model 
\cite{arima77,otsuka78} suggests that 
the number of bosons of each type ($N_{\rho}$) is fixed and is taken as 
the sum of valence proton and neutron particle and hole pairs counted from 
the nearest closed shell. The proton-neutron degrees of freedom are naturally 
reflected in the IBM-2 via an SU(2) F-spin algebra \cite{arima77} 
with generators 
$\hat{F}_{+} = s^{\dagger}_{\pi} s_{\nu} + d^{\dagger}_{\pi}\cdot 
\tilde d_{\nu}$, $\hat{F}_{-} = (\hat{F}_{+})^{\dagger}$, 
$\hat{F}_{0} = (\hat{N}_{\pi} - \hat{N}_{\nu})/2$. 
The basic F-spin doublets are  $(s^{\dagger}_{\pi},s^{\dagger}_{\nu})$, 
and $(d^{\dagger}_{\pi \mu},d^{\dagger}_{\nu \mu})$, 
with F-spin projection +1/2 ($-$1/2) for proton (neutron) bosons.
In a given nucleus, with fixed $N_{\pi}$, $N_{\nu}$, all states have the 
same value of $F_{0}= (N_{\pi} - N_{\nu})/2$, while the allowed 
values of the F-spin quantum number F range from $|F_{0}|$ to 
$F_{max} \equiv (N_{\pi}+N_{\nu})/2 \equiv N/2$ in unit steps. 
F-spin characterizes the $\pi$-$\nu$ symmetry properties of IBM-2 states.
States with maximal F-spin, 
$F \equiv F_{max}$, are fully symmetric and correspond to the IBM-1 
states with only one type of bosons \cite{iac87}. There 
are several arguments, {\it e.g.}, the empirical success of IBM-1, 
the identification of F-spin multiplets [4-7] 
(series of nuclei with constant $F$ and varying $F_{0}$ with nearly 
constant excitation energies), 
and weakness of M1 transitions, 
which lead to the belief that low lying collective states have predominantly 
$F=F_{max}$ \cite{lipas90}. States with $F<F_{max}$, correspond to 
`mixed-symmetry' states, 
most notably, the orbital magnetic dipole scissors mode \cite{boh84} 
has by now been established experimentally as a general 
phenomena in deformed even-even nuclei \cite{rich95}. 

Various procedures have been proposed to estimate the F-spin purity of low 
lying states \cite{lipas90}. These involve exploiting the data on 
M1 transitions 
(which should vanish between pure $F=F_{max}$ states), 
extracting the difference in proton-neutron deformations from 
pion charge exchange \cite{lev90}, 
using ratios of $\gamma$ and ground band magnetic moments \cite{gino92b} and 
the experimental g-factors of $2^{+}_1$ states \cite{wolf93}, and 
considering the excitation energy of mixed symmetry states. 
In the majority of analyses the F-spin admixtures in low 
lying states are found to be of a few percents ($<10\%$), typically 
$2\%-4\%$ \cite{lipas90}. 
In spite of its appeal, however, F-spin cannot be an exact symmetry of the 
Hamiltonian. The assumption of F-spin scalar Hamiltonians is at variance 
with the microscopic interpretation of the IBM-2, 
which necessitates different effective 
interactions between like and unlike nucleons. 
Furthermore, if F-spin was a symmetry of the Hamiltonian, 
then {\it all} states 
would have good F-spin and would be arranged in F-spin multiplets. 
Experimentally this is not case. As noted in an analysis 
\cite{brentano85,gupta89} of rare earth nuclei, the ground bands are in 
F-spin multiplets, whereas the 
vibrational $\beta$ bands and some $\gamma$ bands do not 
form good F-spin multiplets. The empirical situation in the deformed Dy-Os 
region is portrayed in Table I and Fig. 1. From Table I it is seen that, 
for $F>13/2$, the
energies of the $L=2^{+}$ members of the $\gamma$ bands vary fast 
in the multiplet and not always monotonically. The variation in the energies 
of the $\beta$ bands is large and irregular. Thus both microscopic and 
empirical arguments rule out F-spin invariance of the Hamiltonian. 
F-spin can at best be an approximate quantum number which is good only for 
a selected set of states while other states are mixed. We are thus 
confronted with a situation of having `special states' endowed with a good 
symmetry which does not arise from invariance of the Hamiltonian.
These are precisely the characteristics of a ``partial symmetry'' 
for which a non-scalar Hamiltonian produces a subset of special 
(at times solvable) states with good symmetry. Such a symmetry 
notion \cite{PDS} was recently applied to nuclei \cite{lev96}, 
to molecules \cite{ping97} and to the study 
of mixed systems with coexisting regularity and chaos \cite{PDSchaos}. 
Previously determined \cite{lev90} non-F-scalar Hamiltonians were shown 
to have solvable ground bands with good F-spin. It is the purpose of this 
Letter to analyze in detail these Hamiltonians and to show that their 
partial F-spin symmetry reproduces the known systematics 
of ground and excited bands. In particular, we find F-spin multiplets in 
only selected bands, and observe common collective 
signatures for the ground and scissors bands in deformed nuclei, 
{\it e.g.} the same F-spin purity and equal moments of inertia. 
We further test a prediction for the existence of F-spin multiplets of 
scissors states.

The ground band in the IBM-2 is represented by an intrinsic state which 
is a product of a proton condensate and a rotated neutron condensate with 
$N_{\pi}$ and $N_{\nu}$ bosons respectively \cite{intrinsic}. 
It depends on the quadrupole 
deformations, $\beta_{\rho},\gamma_{\rho}$, ($\rho=\pi,\nu$) of the 
proton-neutron equilibrium shapes and on the relative orientation angles 
$\Omega$ between them. For $\beta_{\rho}>0$, the intrinsic state is deformed 
and members of the rotational ground-state band are obtained from it by 
projection. It has been shown in \cite{lev90} that the intrinsic state will 
have a well defined F-spin, $F=F_{max}$, when the proton-neutron shapes are 
aligned and with equal deformations. The conditions ($\beta_{\pi}=
\beta_{\nu}$,$\gamma_{\pi}=\gamma_{\nu}$, $\Omega=0$) are weaker than the 
conditions for F-spin invariance, which makes it possible for a 
non-F-scalar IBM-2 Hamiltonian to have an equilibrium intrinsic state 
with pure F-spin. Since the angular momentum projection 
operator is an F-spin scalar, the projected states of good L will also have 
good $F=F_{max}$. A non-F-spin scalar Hamiltonian which has the 
above equilibrium condensate as an eigenstate is therefore 
guaranteed to have a ground band with good F-spin symmetry. Such explicit 
construction of an IBM-2 Hamiltonian with partial F-spin symmetry was 
presented in \cite{lev90} for the most likely situation, namely, 
aligned axially symmetric (prolate) deformed shapes ($\beta_{\rho}=\beta$, 
$\gamma_{\rho}=\Omega=0$). In this case, the equilibrium deformed 
intrinsic state for the ground band with $F=F_{max}$ has the form
\ba 
\vert c; K=0 \rangle &\equiv& \vert N_{\pi},N_{\nu} \rangle = 
(N_{\pi}!N_{\nu}!)^{- 1/2}(b^{\dagger}_{c,\pi})^{N_{\pi}} \,
(b^{\dagger}_{c,\nu})^{N_{\nu}} \vert 0 \rangle ~,
\nonumber\\
 b^{\dagger}_{c,\rho} &=& (1 + \beta^{2}\,)^{-1/2}
(\,s^{\dagger}_{\rho} + \beta\, d^{\dagger}_{\rho, 0} 
\, ) ~,
\label{cond}
\ea
where $K$ denotes the angular momentum projection on the symmetry axis. 
The relevant IBM-2 Hamiltonian with partial F-spin symmetry can be 
transcribed in the form
\ba
H  &=& 
\sum_{i}\sum_{L=0,2}
A^{(L)}_{i}R^{\dagger}_{i,L}\cdot\tilde{R}_{i,L}
+ \sum_{L=1,2,3}B^{(L)}W^{\dagger}_{L}\cdot\tilde W_{L}
+C^{(2)}\Bigl [\, R^{\dagger}_{(\pi\nu),2}\cdot\tilde W_{2} 
+ H.c.\,\Bigr ] ~, 
\label{hamilt}
\ea
where $H.c.$ means Hermitian conjugate and the dot implies a scalar product.
The $R^{\dagger}_{i,L}$ ($L=0,2$) 
are boson pairs with $F=1$ and 
$(F_0=1,0,-1)\leftrightarrow (i=\pi,(\pi\nu),\nu)$, and $W^{\dagger}_{L}$ 
$(L=1,2,3)$ are F-spin scalar ($F=0$) boson pairs defined as 
\ba
\begin{array}{ll}
R^{\dagger}_{\rho,0} = 
d^{\dagger}_{\rho} \cdot d^{\dagger}_{\rho} -       
\beta ^{2}(s^{\dagger}_{\rho})^2,\;\;\; &
R^{\dagger}_{(\pi\nu),0} = 
\sqrt{2}\,(\,d^{\dagger}_{\pi}\cdot 
d^{\dagger}_{\nu} - \beta^{2}s^{\dagger}_{\pi}s^{\dagger}_{\nu}\,) \\
R^{\dagger}_{\rho,2} = 
\sqrt{2}\,\beta\, s^{\dagger}_{\rho}d^{\dagger}_{\rho} 
+ \sqrt{7}(d^{\dagger}_{\rho}d^{\dagger}_{\rho})^{(2)}, \;\;\;&
R^{\dagger}_{(\pi\nu),2} = 
\beta (\,s^{\dagger}_{\pi}d^{\dagger}_{\nu} 
+ s^{\dagger}_{\nu}d^{\dagger}_{\pi}\,) +                            
\sqrt{14}(d^{\dagger}_{\pi} d^{\dagger}_{\nu})^{(2)} \\
W^{\dagger}_{L} = (d^{\dagger}_{\pi} d^{\dagger}_{\nu})^{(L)}
\;\; (L=1,3),\;\;\; &
W^{\dagger}_{2} = 
s^{\dagger}_{\pi}d^{\dagger}_{\nu} - s^{\dagger}_{\nu}d^{\dagger}_{\pi}
\label{pairs}
\end{array}
\ea
with $\rho=\pi,\nu$ and $\tilde{R}_{i,L,\mu} = (-1)^{\mu}R_{i,L,-\mu}$, 
$\tilde{W}_{L,\mu} = (-1)^{\mu}W_{L,-\mu}$. 
The pair operators satisfy $R_{i,L,\mu}\vert c\rangle = 
W_{L,\mu}\vert c\rangle = 0$ 
and consequently, the condensate is a zero energy 
eigenstate of H for {\it any} choice of parameters 
$A^{(L)}_{i},\,B^{(L)},\, C^{(2)}$ and any {\it any} $N_{\pi},N_{\nu}$.
When $A^{(L)}_{i},\,B^{(L)},\,A^{(2)}_{\pi\nu}B^{(2)}-(C^{(2)})^{2}\geq 0$, 
the above Hamiltonian is positive-definite and hence $\vert c\rangle$ 
is its exact ground state with $F=F_{max}$. 
$H$, however, is an F-spin scalar only 
when $A^{(L)}_{\pi}=A^{(L)}_{\nu}=A^{(L)}_{\pi\nu},\, (L=0,2)$ 
and $C^{(2)}=0$. We thus have a non-F-spin scalar Hamiltonian with a 
solvable (degenerate) ground band with $F=F_{max}$. 
The degeneracy can be lifted 
by adding to the Hamiltonian (F-spin scalar) SO(3) rotation terms which 
produce $L(L+1)$ type splitting but do not affect the wave functions. 
States in other bands can be mixed with respect to F-spin, 
hence the F-spin symmetry of H is partial. $H$ trivially commutes with 
$\hat{F}_{0}$ but not with $\hat{F}_{\pm}$. However, 
$[\,H,\hat{F}_{\pm}\,]\vert c\rangle =0$ 
does hold and therefore $H$ will yield 
F-spin multiplets for members of ground bands. On the other hand, states in 
other bands can have F-spin admixtures and are not compelled to form F-spin 
multiplets. These features which arise from the partial F-spin symmetry 
of the Hamiltonian are in line with the empirical situation as discussed 
above and as depicted in Table I and Fig. 1. 
It should be noted that the partial F-spin symmetry of H holds for any 
choice of parameters in Eq. (\ref{hamilt}). In particular, one can 
incorporate realistic shell-model based constraints, by choosing
the $A^{(2)}_{\rho}$ ($\rho=\pi,\nu$) terms (representing 
seniority-changing interactions between like nucleons), to be small.
For the special choice $A^{(2)}_{i}=C^{(2)}=0$ 
and $B^{(1)}=B^{(3)}$, $H$ of Eq. (\ref{hamilt}) becomes 
SO(5) scalar which commutes, therefore, with the SO(5) projection operator 
and hence produces F-spin multiplets with good SO(5) symmetry. Such 
multiplets were reported in the Yb-Os region of $\gamma$-soft 
nuclei \cite{zamfir92}. 

The same conditions ($\beta_{\rho}=\beta$, $\gamma_{\rho}=\Omega=0$) which 
resulted in $F=F_{max}$  for the condensate of Eq. (\ref{cond}), ensure 
also $F=F_{max}-1$ for the intrinsic state representing the scissors band
\ba
\vert sc ; K=1 \rangle &=& 
\Gamma^{\dagger}_{sc}\vert N_{\pi}-1,N_{\nu}-1 \rangle ~,
\nonumber\\
\Gamma^{\dagger}_{sc} &=& 
b^{\dagger}_{c,\pi}d^{\dagger}_{\nu,1} - 
d^{\dagger}_{\pi,1}b^{\dagger}_{c,\nu} ~.
\label{scissors}
\ea
Here $\Gamma^{\dagger}_{sc}$ is a $F=0$ deformed boson pair whose action 
on the condensate with $(N-2)$ bosons produces the scissors mode 
excitation. Furthermore, the scissors intrinsic state (\ref{scissors}) 
is an exact eigenstate of the following Hamiltonian, obtained from 
Eq. (\ref{hamilt}) for the special choice $C^{(2)} = 0$ and 
$B^{(1)}= B^{(3)} = 2B^{(2)} \equiv 2B$
\ba
H' &=& 
\sum_{i}\sum_{L=0,2}
A^{(L)}_{i}R^{\dagger}_{i,L}\cdot\tilde{R}_{i,L} 
+ B\hat{{\cal M}}_{\pi\nu}
\label{hprime}
\ea
The last term in Eq. (\ref{hprime}) is the Majorana operator 
\cite{iac87}, related to the total F-spin operator by 
${\hat{\cal M}}_{\pi\nu}= [\,\hat {N}(\hat{N}+2)/4 - \hat{F}^2 \,]$, 
with eigenvalues 
$k(N-k+1)$ for states with $F=F_{max}-k$.
The Hamiltonian $H'$ is non-F-scalar but is rotational invariant. 
If we add to it an SO(3) rotation term, $H' +\lambda\hat{L}^2$, 
($\hat{L} = \hat{L}_{\pi}+\hat{L}_{\nu}$), the resulting Hamiltonian
will have a subset of {\it solvable} states which form the $K=0$ ground 
band ($L=0,2,4,\ldots $) with $F=F_{max}$, and the $K=1$ scissors band 
($L=1,2,3\ldots $) with $F=F_{max}-1$. The resulting spectrum is
\ba
\begin{array}{ll}
E_{g}(L) = \lambda\,L(L+1) & (F=F_{max}) \\
E_{sc}(L) = B\,N + \lambda\,L(L+1) \qquad & (F =F_{max} -1)
\end{array}
\label{ene}
\ea
where the Majorana coefficient $B$ may depend on the boson numbers 
and deformation \cite{lipas90,enders99,pietralla98}. 
It follows that for such Hamiltonians, with partial F-spin symmetry, 
both the ground and scissors band have good F-spin and have the same 
moment of inertia. The latter derived property is in agreement with the 
conclusions of a recent comprehensive analysis of the scissors mode in 
heavy even-even nuclei \cite{enders99}, which concluded that, 
within the experimental 
precisions ($\sim$ 10\%), the moment of inertia of the scissors mode are 
the same as that of the ground band. It is the partial F-symmetry of the 
Hamiltonian (\ref{hprime}) which is responsible for the common signatures 
of collectivity in these two bands.

The Hamiltonian $H'$ of Eq. (\ref{hprime}) is not F-spin invariant, 
however, $[\,H' , \vec{F} \,]\,\vert c; K=0 \rangle = 
[\,H' , \vec{F} \,]\,\vert sc; K=1 \rangle = 0$. 
This implies that members of both the ground and scissors bands are 
expected to form F-spin multiplets. For ground bands such structures have 
been empirically established [4-7]. The prediction for F-spin multiplets of 
scissors states requires further elaboration. Although the mean energy of 
the scissors mode is at about 3 MeV \cite{pietralla98}, 
the observed fragmentation of the 
M1 strength among several $1^{+}$ states prohibits, unlike ground bands, 
the use of nearly constant excitation energies as a criteria to 
identify F-spin multiplets of scissors states. 
Instead, a more sensitive test of this suggestion comes from the 
summed ground to scissors B(M1) strength. The IBM-2 
M1 operator $(\hat{L}_{\pi}- \hat{L}_{\nu})$ 
is an F-spin vector ($F=1,F_{0}=0$). Its matrix element between the  
ground state [$L=0^{+}_{g},\,(F = F_{max},F_{0})$] and scissors state 
[$L=1^{+}_{sc},\,(F' =F-1,F_{0}$)] is proportional to an F-spin 
Clebsch Gordan coefficient $C_{F,F_0} = (F,F_0;1,0\vert F-1,F_0)$ times 
a reduced matrix element. It follows that the ratio  
$B(M1;0^{+}_{g}\rightarrow 1^{+}_{sc})/(C_{F,F_0})^2$ does not 
depend on $F_{0}$ and should be a constant in a given F-spin multiplet. 
In Table II we list 
{\it all} F-spin partners for which the summed B(M1) strength 
to the scissors mode has been measured todate \cite{pietralla95,maser96}. 
It is seen that within the experimental errors, the above ratio is fairly 
constant. The most noticeable discrepancy for $^{172}$Yb (F=8), arises 
from its measured low value of summed B(M1) strength. 
The latter should be regarded as a lower limit due to experimental 
deficiencies (large background and strong fragmentation \cite{rich95}).
These observations strengthen the contention of high F-spin purity and 
formation of F-spin multiplets of scissors states.

As noted in \cite{brentano85,gupta89} and shown in Table I and Fig. 1, for 
nuclei with $F=6,\, 6.5,\,$ also members of the $\gamma$ bands display 
constant excitation energies and seem to form good F-spin multiplets. 
This empirical observation has a natural explanation within the family of 
Hamiltonians with partial F-spin symmetry. 
For the choice $\beta=\sqrt{2}$ and 
$A^{(2)}_{\pi}=A^{(2)}_{\nu}=A^{(2)}_{\pi\nu}$ 
in Eq. (\ref{hprime}), $H'$ will have both F-spin 
and SU(3) partial symmetries. In such circumstances, the ground ($K=0$), 
scissors ($K=1$), symmetric-$\gamma$ ($K=2$), and antisymmetric-$\gamma$ 
($K=2$) bands are solvable and have good SU(3) and F-spin symmetries: 
$[(\lambda,\mu),F] = [(2N,0),F_{max}]$, $[(2N-2,1),F=F_{max}-1]$, 
$[(2N-4,2),F=F_{max}]$ and $[(2N-4,2),F=F_{max}-1]$ respectively.
The intrinsic states for the symmetric-$\gamma$ or antisymmetric-$\gamma$ 
bands are obtained by F-spin coupling the $F=1$ pair 
$R^{\dagger}_{i,2,\mu=2}$ to
the ($F=F_{max}-1$) condensate $\vert N_{\pi}-1,N_{\nu}-1\rangle$ 
with ($N-2$) bosons to form a N-boson intrinsic state with 
$F=F_{max}$ or $F=F_{max}-1$. Since, in this case, the commutator 
$[\,H' ,\vec{F}\,]$ vanishes when it acts on the 
solvable intrinsic states,
the projected states are ensured to have good F-spin and form 
F-spin multiplets. At the same time, since the Hamiltonian is not F-spin 
scalar, the $\beta$ bands can have F-spin admixtures and need not form 
F-spin multiplets. 
 
In summary, we have examined in detail IBM-2 Hamiltonians with partial 
F-spin symmetry. The latter are not F-spin scalars, 
yet have a subset of solvable eigenstates with good F-spin symmetry. 
In particular, the corresponding ground bands form F-spin multiplets 
with $F=F_{max}$, but excited bands can be mixed, 
which is in line with the empirically observed F-spin multiplets [4-7]. 
A class of IBM-2 Hamiltonians with partial F-spin 
symmetry predict the occurrence of F-spin multiplets of scissors states, 
with a moment of inertia equal to that of the ground band. 
This prediction is in agreement with recent analyses of the 
empirical systematics of excitation energy and M1 strength of the scissors 
mode in even-even nuclei \cite{enders99,pietralla98}. 
All the above findings illuminate the potential useful role of F-spin 
(and other) partial symmetries in nuclear spectroscopy and motivate their 
further study.

We acknowledge helpful discussions with N. Pietralla on the empirical data 
and thank the Institute for Nuclear Theory at the University of Washington 
for its hospitality. This work was was supported in part by the Israel 
Science Foundation, the Zevi Hermann Schapira research fund, and by the 
U.S. Department of Energy under contract W-7405-ENG-36.

\pagebreak

\newpage
\begin{table}
\caption[]{\small 
Energies (in MeV) of $2^{+}$ levels of the ground ($g$), 
$\gamma$ and $\beta$ bands in F-spin multiplets. The mass numbers are 
$A= 132 + 4F$.
\normalsize}
\vskip 10pt
\begin{tabular}{lccccccc}
F & Energy & $^{A}$Dy & $^{A+4}$Er & $^{A+8}$Yb & $^{A+12}$Hf & 
 $^{A+16}$W & $^{A+20}$Os \\
\hline
6   & $E(2^{+}_{g})$      & 0.14 & 0.13 & 0.12 & 0.12 & 0.12 & 0.14 \\
    & $E(2^{+}_{\gamma})$ & 0.89 & 0.85 & 0.86 & 0.88 &      & 0.86 \\
    & $E(2^{+}_{\beta})$  & 0.83 & 1.01 & 1.07 & 1.06 &      & 0.74 \\
\hline
13/2 & $E(2^{+}_{g})$     & 0.10 & 0.10 & 0.10 & 0.10 & 0.11 & 0.13 \\ 
    & $E(2^{+}_{\gamma})$ & 0.95 & 0.90 & 0.93 & 0.96 &      &      \\
    & $E(2^{+}_{\beta})$  & 1.09 & 1.17 & 1.14 & 0.99 &      &      \\
\hline
7   & $E(2^{+}_{g})$      & 0.09 & 0.09 & 0.09 & 0.10 & 0.11 & 0.13 \\
    & $E(2^{+}_{\gamma})$ & 0.97 & 0.86 & 0.98 & 1.08 &      & 0.87 \\
    & $E(2^{+}_{\beta})$  & 1.35 & 1.31 & 1.23 & 0.95 &      & 0.83 \\
\hline
15/2 & $E(2^{+}_{g})$     & 0.08 & 0.08 & 0.08 & 0.09 & 0.11 &      \\
    & $E(2^{+}_{\gamma})$ & 0.89 & 0.79 & 1.15 & 1.23 & 1.11 &      \\
    & $E(2^{+}_{\beta})$  & 1.45 & 1.53 & 1.14 & 0.90 & 1.08 &      \\
\hline
8   & $E(2^{+}_{g})$      & 0.07 & 0.08 & 0.08 & 0.09 &      &      \\  

    & $E(2^{+}_{\gamma})$ & 0.76 & 0.82 & 1.47 & 1.34 &      &      \\
    & $E(2^{+}_{\beta})$  &      & 1.28 & 1.12 & 1.23 &      &      \\
\hline
17/2 & $E(2^{+}_{g})$     & 0.08 & 0.08 & 0.08 &      &      &      \\
    & $E(2^{+}_{\gamma})$ & 0.86 & 0.93 & 1.63 &      &      &      \\
    & $E(2^{+}_{\beta})$  & 1.21 & 0.96 & 1.56 &      &      &      \\
\end{tabular}
\end{table}

\begin{table}[]
\centering
\caption[]{\small 
The ratio $R=\sum B(M1)\uparrow/(C_{F,F_0})^2$ 
for members of F-spin~multiplets. 
Here $\sum B(M1)\uparrow$ denotes summed M1 strength to the scissors mode 
and $C_{F,F_0} = (F,F_0;1,0\vert F-1,F_0)$.
Data taken from \cite{pietralla95,maser96}.
\normalsize}
\vskip 10pt
\begin{tabular}{lccccc}
Nucleus & $F$ & $F_0$ & $\sum B(M1)\uparrow$ $[\mu_{N}^2]$ 
& $(C_{F,F_0})^2$ & $R$ \\
\hline
$^{148}$Nd & 4    & 1    & 0.78 (0.07) & 5/12     & 1.87 (0.17) \\
$^{148}$Sm &      & 2    & 0.43 (0.12) & 1/3      & 1.29 (0.36) \\
\hline
$^{150}$Nd & 9/2  & 1/2  & 1.61 (0.09) & 4/9      & 3.62 (0.20) \\
$^{150}$Sm &      & 3/2  & 0.92 (0.06) & 2/5      & 2.30 (0.15) \\
\hline
$^{154}$Sm & 11/2 & 1/2  & 2.18 (0.12) & 5/11     & 4.80 (0.26) \\
$^{154}$Gd &      & 3/2  & 2.60 (0.50) & 14/33    & 6.13 (1.18) \\
\hline
$^{160}$Gd & 7    & 0    & 2.97 (0.12) & 7/15     & 6.36 (0.26) \\
$^{160}$Dy &      & 1    & 2.42 (0.18) & 16/35    & 5.29 (0.39) \\
\hline
$^{162}$Dy & 15/2 & 1/2  & 2.49 (0.13) & 7/15     & 5.34 (0.28) \\
$^{166}$Er &      & $-1/2$ & 2.67 (0.19) & 7/15   & 5.72 (0.41) \\
\hline
$^{164}$Dy & 8    & 0    & 3.18 (0.15) & 8/17     & 6.76 (0.32) \\
$^{168}$Er &      & $-1$   & 3.30 (0.12) & 63/136 & 7.12 (0.26) \\
$^{172}$Yb &      & $-2$   & 1.94 (0.22)$^{a)}$ & 15/34 & 4.40 (0.50) \\
\hline
$^{170}$Er & 17/2 & $-3/2$ & 2.63 (0.16) & 70/153 & 5.75 (0.35) \\
$^{174}$Yb &      & $-5/2$ & 2.70 (0.31) & 66/153 & 6.26 (0.72) \\
\end{tabular}
{\small $^{a)}$ The low value of $\sum B(M1)\uparrow $ 
for $^{172}$Yb has been attributed to experimental deficiencies 
\cite{rich95}.
$\qquad\qquad$}\\
\end{table}

\clearpage
\begin{figure}
\caption{Experimental levels of the ground $\gamma$ and $\beta$ bands 
in an F-spin multiplet $F=6$ of rare earth nuclei. 
Levels shown are up to $L=8^{+}_{g}$ for the 
ground band, $L=2^{+}_{\gamma},3^{+}_{\gamma}$ for the $\gamma$ band 
(diamonds connected by dashed lines)
and $L=0^{+}_{\beta},2^{+}_{\beta}$ for the $\beta$ band 
(squares connected by dotted lines).}
\end{figure}


\begin{thebibliography}{99}

\bibitem{iac87}
F. Iachello and A. Arima, {\it The Interacting Boson Model} (Cambridge
University Press, Cambridge, 1987).

\bibitem{arima77}
A. Arima, T. Otsuka, F. Iachello and I. Talmi, 
Phys. Lett. B {\bf 66}, 205 (1977).

\bibitem{otsuka78}
T.~Otsuka, A.~Arima, F.~Iachello and I.~Talmi, 
Phys. Lett. B {\bf 76}, 139 (1978). 

\bibitem{harter85}
H. Harter, P. von Brentano, A. Gelberg and R.F. Casten,
Phys. Rev. C {\bf 32}, 631 (1985).

\bibitem{brentano85}
P. von Brentano, A. Gelberg, H. Harter and P. Sala,
J. Phys. G {\bf 11}, L85 (1985).

\bibitem{gupta89}
J.B. Gupta,
Phys. Rev. C {\bf 39}, 272 (1989).

\bibitem{zamfir92}
N.V. Zamfir, R.F. Casten, P. von Brentano and W.-T. Chou, 
Phys. Rev. C {\bf 46}, R393 (1992).

\bibitem{lipas90}
P.O. Lipas, P. von Brentano and A. Gelberg,
Rep. Prog. Phys. {\bf 53}, 1355 (1990) and references therein.

\bibitem{boh84}
D. Bohle {\it  et al.}, Phys. Lett. {\bf B137}, 27 (1984).

\bibitem{rich95}
For recent reviews see, 
A. Richter, Prog. Part. Nucl. Phys. {\bf 34}, 261 (1995);
U. Kneissl, H.H. Pitz and A. Zilges, 
Prog. Part. Nucl. Phys. {\bf 37}, 349 (1996).

\bibitem{lev90}
A. Leviatan, J.N. Ginocchio and M.W. Kirson,
Phys. Rev. Lett. {\bf 65}, 2853 (1990).

\bibitem{gino92b}
J.N. Ginocchio, W. Frank and P. von Brentano,
Nucl. Phys. A {\bf 541}, 211 (1992).

\bibitem{wolf93}
A Wolf, O. Scholten and R.F. Casten, 
Phys. Lett. B {\bf 312}, 372 (1993).

\bibitem{PDS} 
Y.~Alhassid and A.~Leviatan, J.~Phys.~A {\bf 25}, L1265 (1992).

\bibitem{lev96} 
A.~Leviatan, Phys.~Rev.~Lett. {\bf 77}, 818 (1996); 
A. Leviatan and I. Sinai J. Phys. G {\bf 25}, 791 (1999).  

\bibitem{ping97}
J.L. Ping and J.Q. Chen,
Ann. Phys. {\bf 255}, 75 (1997).

\bibitem{PDSchaos} 
N.~D.~Whelan, Y.~Alhassid and A.~Leviatan,
Phys. Rev. Lett. {\bf 71}, 2208 (1993);
A.~Leviatan and N.~D.~Whelan, Phys. Rev. Lett. {\bf 77}, 5202 (1996).

\bibitem{intrinsic}
J.N. Ginocchio and M.W. Kirson, Nucl. Phys. A {\bf 350}, 31 (1980); 
A. Leviatan and M. W. Kirson, Ann. Phys. {\bf 201}, 13 (1990). 

\bibitem{enders99}
J. Enders {\it et al.}, 
Phys. Rev. C {\bf 59}, R1851 (1999).

\bibitem{pietralla98}
N. Pietralla {\it et al.},
Phys. Rev. C {\bf 58}, 184 (1998).

\bibitem{pietralla95}
N. Pietralla {\it et al.}, 
Phys. Rev. C {\bf 52}, R2317 (1995) and references therein.

\bibitem{maser96}
H. Maser {\it et al.}, 
Phys. Rev. C {\bf 53}, 2749 (1996).

\end{thebibliography}
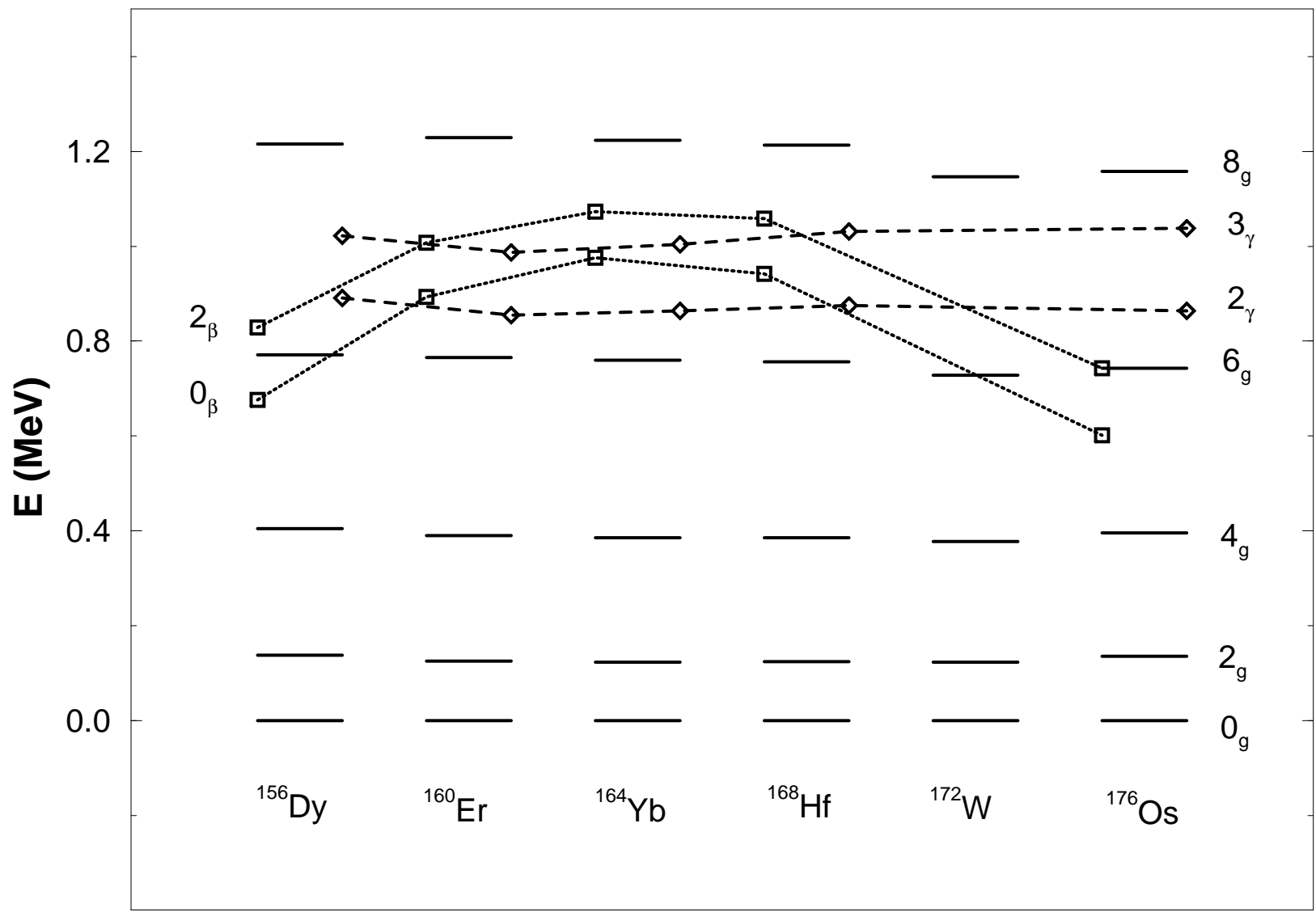
\end{document}